\documentclass[%
 reprint,
 showpacs,
 nofootinbib,
 amsmath,amssymb,
 aps,
 pra,
]{revtex4-1}

\usepackage{graphics}
\usepackage{graphicx}
\usepackage{color}
\usepackage{amsmath}
\usepackage{slashed}
\usepackage{epsfig}
\usepackage[hyperindex,breaklinks]{hyperref}
\PassOptionsToPackage{linktocpage}{hyperref}

\begin{document}

\title{Phenomenology of a very light scalar (100~MeV~$<m_h<$~10~GeV) \\ mixing with the SM Higgs}

\author{Jackson D. Clarke, Robert Foot, Raymond R. Volkas}
\affiliation{ARC Centre of Excellence for Particle Physics at the Terascale, \\ 
School of Physics, University of Melbourne, VIC 3010, Australia.}

\date{\today}

\begin{abstract}

In this paper we investigate the phenomenology of a very light scalar, $h$, with mass 100~MeV~$<m_h<$~10~GeV, mixing with the SM Higgs. As a benchmark model we take the real singlet scalar extension of the SM. We point out apparently unresolved uncertainties in the branching ratios and lifetime of $h$ in a crucial region of parameter space for LHC phenomenology. Bounds from LEP, meson decays and fixed target experiments are reviewed. We also examine prospects at the LHC. For $m_h \lesssim m_B$ the dominant production mechanism is via meson decay; our main result is the calculation of the differential $p_T$ spectrum of $h$ scalars originating from B mesons and the subsequent prediction of up to thousands of moderate (triggerable) $p_T$ displaced dimuons possibly hiding in the existing dataset at ATLAS/CMS or at LHCb. We also demonstrate that the subdominant $Vh$ production channel has the best sensitivity for $m_h \gtrsim m_B$ and that 
future bounds in this region could conceivably compete with those of LEP.

\end{abstract}


\maketitle

\section{Introduction}

The recent discovery of a resonance at mass $\sim 125$~GeV \cite{Aad:2012tfa,*Chatrchyan:2012ufa}, with properties consistent 
with those of the standard model (SM) Higgs boson, $H$,
appears to confirm the basic picture of electroweak symmetry breaking. That is, $SU(2)\otimes U(1)$ gauge
symmetry is spontaneously broken by the nontrivial vacuum of an elementary scalar field.
An important question arises: Are there any more elementary scalars? 

Additional scalars are required to exist in various extensions of the SM.
In particular, realistic perturbative Coleman-Weinberg \cite{PhysRevD.7.1888} models with classical scale invariance broken radiatively and spontaneously 
can be constructed \cite{Hempfling:1996ht,*Meissner:2006zh,*Chang:2007ki,*Foot:2007as,*Foot:2007ay,*Foot:2007iy,*Iso:2009ss,*Iso:2009nw,*Holthausen:2009uc,*Foot:2010av,*AlexanderNunneley:2010nw,*Ishiwata:2011aa,*Lee:2012jn,*Okada:2012sg,*Iso:2012jn,*Englert:2013gz,*Heikinheimo:2013fta,*Heikinheimo:2013xua,*Hambye:2013dgv,*Bars:2013yba,*Heikinheimo:2013cua,*Carone:2013wla,*Farzinnia:2013pga,*Khoze:2013uia,*Gabrielli:2013hma,*Antipin:2013exa}.
Such models generally feature at least one additional (real) singlet scalar, $S$. If scale invariance is
broken at the electroweak scale, by the VEV $\langle S \rangle$, then a GeV-scale scalar state, $h$, 
is predicted \cite{Foot:2011et}. This state is the pseudo-Goldstone boson associated with the spontaneous breaking
of scale invariance \cite{PhysRevD.13.3333}.  The mass eigenstates, $h$ and $H$, are, in general, an orthogonal rotation
of the weak eigenstates: 
\begin{align}
 \left(
 \begin{array}{c}
  H \\
  h
 \end{array}
 \right)
 =
 \left(
 \begin{array}{cc}
  \cos\rho 	& -\sin\rho \\
  \sin\rho 	& \cos\rho
 \end{array}
 \right)
  \left(
 \begin{array}{c}
  \phi'_0 \\
  S'
 \end{array}
 \right),
\end{align}
where $\phi'_0$ is a pure doublet component and $\rho$ is a mixing angle.  

In scale invariant models, the Cosmological Constant (CC) is a finite and calculable parameter.
Setting it to be small, consistent with observations, leads to non-trivial constraints on the 
parameters of the theory \cite{Foot:2010et}. Applied to electroweak scale invariant models with a real singlet scalar, the CC
constraint implies \cite{Foot:2011et} that the effective couplings $Hhh$ and $Hhhh$ are very small and
the mass of $h$ and the angle $\rho$ are correlated: 
\begin{align}
 \sin^2\rho &\sim \frac{m_h^2}{500\text{ GeV}^2}\ .
\end{align}
We refer to this throughout as the Foot \& Kobakhidze prediction.

There are also other quite different motivations for being interested in light scalars.
Bezrukov \& Gorbunov \cite{Bezrukov:2009yw,Bezrukov:2013fca} have considered a class of inflationary models which
feature a light scalar; constraints from primordial density perturbations imply the relation
\begin{align}
 \sin^2\rho &\sim \frac{2\times 10^{-8}\text{ GeV}^2}{m_h^2} \ .
\end{align}

More generally, some hidden sector (which may or may not contain dark matter) might exist which couples to the singlet scalar.
In this case the so-called Higgs portal quartic interaction term then facilitates interactions involving the two sectors.
Depending on the mass of the hidden states, invisible decays of $h$ and/or $H$ could be allowed. This occurs, for example, 
in the recent model of Weinberg motivated by hints of a fractional cosmic neutrino excess \cite{Weinberg:2013kea} 
(see also Refs.~\cite{Cheung:2013oya,*Garcia-Cely:2013nin} for some phenomenological analyses). Another possibility is that hidden states decay 
back into SM particles on collider length scales after production, possibly resulting in distinctive signatures such as displaced vertices and/or
high multiplicity cascade decays like those seen in hidden valley models \cite{Strassler:2006im}.

The purpose of this paper is to survey the phenomenological consequences of a very light scalar, with mass 100~MeV~$<m_h<$~10~GeV, mixing with the SM Higgs.
As a benchmark model we take the real singlet scalar extension of the SM. In this case, $h$ decays only to SM particles with a vertex factor $\sin \rho$ compared to the SM Higgs. 
The production cross section in all channels we consider is proportional to $\sin^2\rho$, the branching fractions are independent of $\sin^2\rho$, 
and the lifetime is inversely proportional to $\sin^2\rho$:
\begin{align}
 c\tau = \frac{c\tau_{SM}}{\sin^2\rho}, 
 \label{eqctau}
\end{align}
where $c\tau_{SM}$ is the mean decay length of a scalar of mass $m_h$ with exactly SM Higgs couplings, i.e. $h$ when $\sin^2\rho=1$. 
Our approach is to explore $(m_h,\sin^2\rho)$ parameter space, which allows us to test the models of Foot \& Kobakhidze and Bezrukov \& Gorbunov concurrently.

In models where $h$ decays also into invisible exotic states,
one may repeat our analysis in the following way: the production cross section is unaffected, 
the branching fraction to SM final states is altered by a (generally) mass-dependent quantity $\mathcal{B}_{SM} \equiv Br(h\to X_{SM})$, 
and the lifetime becomes shorter by a factor $\mathcal{B}_{SM}$. One would also need to take into account the branching to 
invisible states for the invisible searches considered. We take $\mathcal{B}_{SM}=1$ in our benchmark model and comment on the $\mathcal{B}_{SM}<1$ case when appropriate.

The paper is organised as follows. In Sec.~\ref{secproperties} we review the properties of our benchmark scalar, discussing some large, 
apparently unresolved uncertainties in branching fractions and lifetime. In Sec.~\ref{secbounds} we determine existing bounds from LEP, 
meson decays, and fixed target experiments. In Sec.~\ref{seclhc} we explore phenomenology and prospects at the LHC. 
Our main result is the prediction of many inclusive displaced dimuon events for $m_h\lesssim m_B$ and the observation that the subdominant 
$Vh$ channel has the best sensitivity for $m_h\gtrsim m_B$. We conclude in Sec.~\ref{secconclusion}.

\section{Properties \label{secproperties}}

Of interest is the value of $Br(h\rightarrow \mu^+\mu^-)$ and the mean decay length of $h$. For $\sin^2\rho=1$, $h$ is a hypothetical SM Higgs boson of mass $m_h$. We may therefore appeal to the literature on the SM Higgs before it was ruled out below $2m_b$ \cite{Gunion:1989we}. 

\begin{figure}[t]
 \includegraphics[width=7cm]{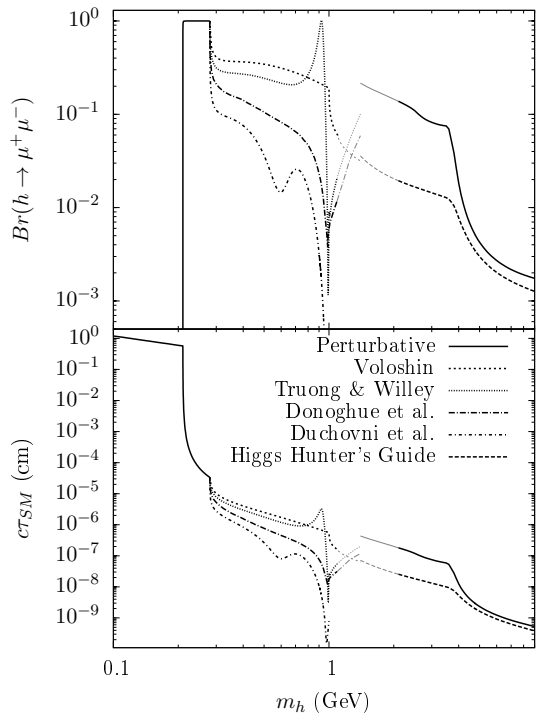}
 \caption{Branching fraction for a light scalar $h$ decaying into muons and its mean decay length for $\sin^2\rho=1$ (see Eq.~\ref{eqctau}) as predicted by a number of models (see text) \cite{Voloshin:1985tc,Duchovni:1989ii,Truong:1989my,Donoghue:1990xh,Gunion:1989we}. The Duchovni et al. prediction is an application of the Raby \& West result \cite{Raby:1988qf}.}
 \label{figbrctau}
\end{figure}

The width to leptons is given by
\begin{align}
 \Gamma(h\rightarrow l^+l^-) = \sin^2\rho\times \frac{m_l^2 m_h}{8\pi v^2}\beta_l^3,
\end{align}
where $\beta_l=\sqrt{1-4m_l^2/m_h^2}$ and $v\approx 246$~GeV. For $m_h<2m_\mu\approx 210$~MeV, $h$ decays almost entirely to $e^+e^-$. Above $2m_\mu$ the decay to $\mu^+\mu^-$ takes over until the $2m_{\pi}\approx 280$~MeV threshold, where the ratio $R_{\pi\mu}=\Gamma(h\rightarrow \pi\pi)/\Gamma(h\rightarrow \mu\mu)$ was historically the subject of much debate \cite{Gunion:1989we,Voloshin:1985tc,Grinstein:1988yu,Raby:1988qf,Duchovni:1989ii,Narison:1989az,Truong:1989my,Donoghue:1990xh}. In Fig.~\ref{figbrctau} we reproduce a selection of results to illustrate the large uncertainty in this mass range attributable to resonant $\pi\pi$ enhancements. We note that Ref.~\cite{Donoghue:1990xh} is the most recent paper, that we are aware of, that is dedicated to the subject. Above the $2m_K\approx 1$~GeV threshold the decay to $KK$ must be taken into account, and has been by a selection of these authors \cite{Narison:1989az,Truong:1989my,Donoghue:1990xh}. Above the $2m_\eta \approx 1.1$~GeV threshold we know of no 
reliable 
prediction. Somewhere 
above 2~GeV, where the energy involved in the decay is much larger than the typical quark binding energy, the perturbative spectator 
approach may be utilised \cite{Gunion:1989we}:
\begin{align}
 \Gamma_{\mu\mu} : \Gamma_{s\bar{s}}& : \Gamma_{c\bar{c}} : \Gamma_{\tau\tau} : \Gamma_{gg} \approx m_\mu^2\beta_\mu^3 : 3m_s^2\beta_K^3 : 3m_c^2\beta_D^3 \nonumber \\ 
 & : m_\tau^2\beta_\tau^3 : \frac{\alpha_s(m_h)^2m_h^2}{9\pi^2}\left| \sum_q I\left( \frac{m_q^2}{m_h^2} \right) \right|^2,
 \label{eqperturbative}
\end{align}
where
\begin{align}
 I(z)=3\left[2z+2z(1-4z)\left(\sin^{-1}\frac{1}{\sqrt{4z}}\right)^2\right].
\end{align}
In Fig.~\ref{figbrctau} we plot this result alongside that of Ref.~\cite{Gunion:1989we}.\footnote{Ref.~\cite{Gunion:1989we} set $m_u=m_d=40$~MeV, $m_s=450$~MeV and $\alpha_s=0.15\pi$ in order to match the result of Ref.~\cite{Voloshin:1985tc} at $m_h\approx 1.5$~GeV; this is no longer well-motivated. We use $m_s=100$~MeV and run $\alpha_s$ according to Figure~17 of Ref.~\cite{Bethke:2006ac}}

The large uncertainties between $2m_\pi<m_h<4$~GeV are apparently unresolved. It would be interesting to know whether a more sophisticated approach is now possible which would provide new insight. A new result would be useful since, in this region, the mean decay length plays an important role in LHC phenomenology.

\section{Bounds \label{secbounds}}

\subsection{LEP}

Constraints from the LEP collider experiment arise from the Bjorken process $e^+e^-\to Z \to Z^\ast h$.

Below $m_h=2m_\mu$, the unboosted mean decay length of $h$ is $\sim 1$~cm~$/\sin^2\rho$. 
With a typical momentum of $\sim8$~GeV \cite{Gross:1992vq} at this mass scale, $h$ escapes the LEP detector and the appropriate bound to apply is that for an invisibly decaying Higgs boson. The $95\%$ C.L. bound is $\sin^2\rho \lesssim 2\times 10^{-3}$ \cite{Acciarri:1996um,Buskulic:1993gi}. The limits given in Refs.~\cite{Acciarri:1996um,Buskulic:1993gi,Abbiendi:2007ac,Searches:2001ab} would also apply to scalars with $\mathcal{B}_{SM}<1$ for $m_h>2m_\mu$.

\begin{figure}
 \includegraphics[width=8cm]{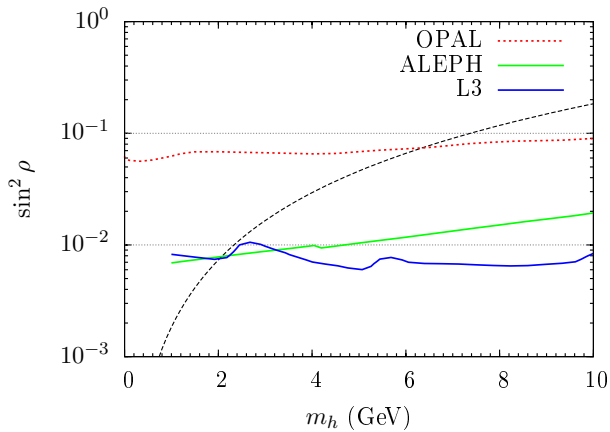}
 \caption{$95\%$ C.L. upper bounds on $\sin^2\rho$ as a function of $m_h$ from OPAL, ALEPH and L3. Also shown is the Foot \& Kobakhidze prediction (dashed).}
 \label{figLEPlims}
\end{figure}

Somewhere not far above $m_h=2m_\mu$, prompt searches become relevant.\footnote{We do not labour on exactly when this occurs, since we find that meson decays set the best limits for $m_h<(m_B-m_K)$.} The best constraints are from the LEP1 searches of ALEPH and L3 \cite{Buskulic:1993gi,Acciarri:1996um}. The $95\%$ C.L. bounds are reproduced in Fig.~\ref{figLEPlims}. For reference we also show the bound from the decay-mode independent search of OPAL using the full LEP1+2 dataset \cite{Abbiendi:2002qp}, which applies to any light scalar regardless of lifetime, branching fractions or exotic decays. These bounds are the best available for $m_h>(m_B-m_K)\approx 4.8$~GeV. They rule out the Foot \& Kobakhidze model for $m_h>2$~GeV.

A short note on these results. L3 considered only hadronic $h$ decays in the $hZ^\ast \to h\nu\nu, h e^+e^-, h\mu^+\mu^-$ channels for $m_h>4$~GeV. ALEPH used $hZ^\ast \to h\nu\nu$, with $h\to hadronic$ or $h\to$ \textit{two/four charged prongs} in the region $2m_\mu<m_h<2m_b$. Figure~5.5 in Ref.~\cite{Decamp:1991uy} shows that, for $m_h>5$~GeV, the efficiency of the charged prong search falls and the hadronic search dominates. Therefore, in this region, the LEP limits are unique in that the limit is set by the hadronic decay of $h$. This is attributable to the comparatively low hadronic background at an $e^+e^-$ collider and the fact that the hadrons appear as a monojet due to the boost of $h$ when $m_h\lesssim 15$~GeV.

We note that LEP limits for $m_h<2m_b$ could have been significantly improved beyond LEP1. The L3 search analysed 114~pb$^{-1}$ of data; the full LEP dataset is $\sim 3000$~pb$^{-1}$. With $\sqrt{s}>(m_Z+m_h)$, production of a real $Zh$ pair becomes significant and background falls away \cite{Sopczak:1994hd}. Instead, analyses focused on the search for the SM Higgs above the $b\bar{b}$ threshold \cite{Barate:2003sz}. We can only surmise that, without motivation, this area of parameter space was overlooked.

\subsection{Meson decays \label{subsecmesons}}

\begin{figure}
 \includegraphics[width=6cm]{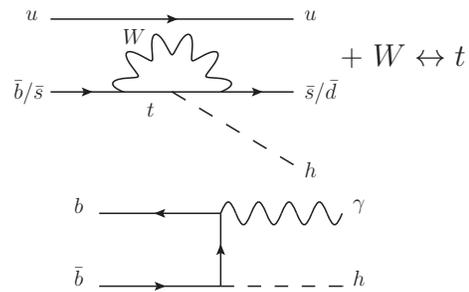}
 \caption{Kaon, B meson, and radiative $\Upsilon$ decays involving $h$.}
 \label{figmesondecays}
\end{figure}

The effective $\bar{s}dh$ ($\bar{b}sh$) vertex contributing to kaon (B meson) decay is obtained by integrating out the top-W loop from the diagram shown in Fig.~\ref{figmesondecays}. This effective vertex leads to the decays $K \rightarrow \pi h \rightarrow \pi \mu^+\mu^-$ and $B \rightarrow K h \rightarrow K \mu^+\mu^-$, with branchings \cite{Leutwyler:1989xj,Batell:2009jf}
\begin{align}
 Br(K^+\to \pi^+ h) 	&\approx \sin^2\rho \times 0.002 \times \frac{2|\vec{p}_h|}{m_K}, \\
 Br(B^+ \to K^+ h) 	&\approx \sin^2\rho \times 0.5 \times \frac{2|\vec{p}_h|}{m_B}\times \mathcal{F}_K^2(m_h),
\end{align}
where $|\vec{p}_h|$ is found using two-body kinematics and the form factor
$\mathcal{F}_K^2(m_h)=\left(1-m_h^2/38\text{ GeV}^2 \right)^{-1}$ \cite{Ball:2004ye}.

In applying experimental constraints from these decays one must properly take into account the lifetime of $h$; either $h$ decays ``promptly enough'' so that the muons are reconstructed with the associated meson, or it does not and the experiment sees missing momentum. In the following, we take into account lifetime by requiring $h$ to decay within a certain (experiment-dependent) distance of the meson decay. For simplicity, and because we only expect a small correction, we do not impose any angular constraints. We stress that, where lifetime has an effect, these can only be considered order of magnitude estimates.

As discussed in Sec.~\ref{secproperties}, there is large uncertainty in the lifetime of $h$ above the $\pi\pi$ threshold. We find that the dependence of the following bounds on $h$ lifetime above this threshold is small, and certainly negligible for $m_h>400$~MeV with the existing experimental reach. We therefore present results as bounds on $\sin^2\rho$ assuming the model of Ref.~\cite{Donoghue:1990xh} below 400~MeV, and unambiguously on $\sin^2\rho\times Br(h\to l^+l^-)$ above, where $l$ corresponds to either $\mu$ or $\tau$, depending on the channel.

\subsubsection{Kaon decays}

\begin{figure}
 \includegraphics[width=8cm]{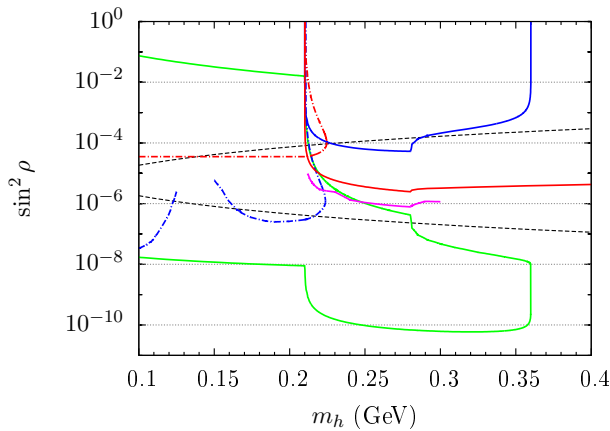}
 \caption{Limits on $(m_h,\sin^2\rho)$ parameter space from meson decays for $m_h<400$~MeV: $K\to \pi\mu^+\mu^-$ (blue solid), $K\to \pi+invisible$ (blue dot-dashed), $B\to K\mu^+\mu^-$ (red solid), $B\to K+invisible$ (red dot-dashed), $B\to K^{\ast0}\mu^+\mu^-$ dedicated search (magenta), and the CHARM beam dump experiment (green enclosed is excluded). Also shown are the predictions from the models of Foot \& Kobakhidze and Bezrukov \& Gorbunov descending (dashed).}
 \label{figmesonlims}
\end{figure}

The NA48/2 collaboration has measured $Br(K^\pm \rightarrow \pi^\pm \mu^+\mu^-)=(9.62 \pm 0.25)\times 10^{-8}$ \cite{Batley:2011zz}, in good agreement with the theoretical predictions $(8.7\pm2.8)\times10^{-8}$ and $(12\pm3)\times10^{-8}$ \cite{Friot:2004yr,*Dubnickova:2006mk}. To derive limits on $\sin^2\rho$ we assume that a $\pi\mu\mu$ vertex is reconstructed if the $h\rightarrow \mu^+\mu^-$ decay occurs within the longitudinal vertex resolution, $\sigma_z\approx 100$~cm \cite{NA482}, of the kaon decay, and not reconstructed otherwise. 
A conservative limit on additive new physics is obtained by taking the difference between the low end of SM theoretical predictions, $Br(K^\pm \rightarrow \pi^\pm \mu^+\mu^-)_{theory}\gtrsim6\times10^{-8}$, and the experimental measurement:
\begin{align}
 Br(K\rightarrow \pi h)&\times Br(h\rightarrow \mu^+\mu^-) \nonumber \\
 &\times \left(1-\exp\left[ \frac{-\sigma_z}{\gamma\beta c\tau} \right]\right) 
 \lesssim 4\times 10^{-8},
\end{align}
where the bracketed term is the probability that a particle with lifetime $\tau$, speed $\beta c$ and boost $\gamma$ decays within a distance $\sigma_z$, and $\gamma\beta\approx 120$ is inherited from the kaon with momentum 60~GeV. Note that both $Br(K\rightarrow \pi h)$ and $c\tau$ depend on $\sin^2\rho$, so that this inequality may be used to constrain $\sin^2\rho$. The obtained constraint is given by the solid blue curve in Fig.~\ref{figmesonlims}.

The E949 collaboration has published a $90\%$ C.L. upper limit on the two-body decay $Br(K^\pm \rightarrow \pi^\pm X)\times Br(X\rightarrow invisible)$ that is better than $10^{-9}$ between 170~MeV and 240~MeV \cite{Artamonov:2009sz}. The limit was derived assuming the decay of $X$ was detected and vetoed with $100\%$ efficiency if $X$ decayed within the outer radius of the barrel veto, $l_{BV}\approx 145$~cm \cite{Adler:2008zza}. We therefore impose the following:
\begin{align}
 Br&(K\rightarrow \pi h) \nonumber \\
 &\times \int_{0}^{\pi}\frac{\sin\theta d\theta}{2}
 \exp\left[ \frac{-l_{BV}}{\sin\theta} \frac{1}{\gamma\beta c\tau} \right] 
 < \text{E949 limit},
\end{align}
where $\gamma\beta\sim 1$ is determined using two-body kinematics assuming a stationary kaon. This bound applies where $h$ escapes the detector; it also applies to invisibly decaying scalars if $\mathcal{B}_{SM}<1$. It is shown as the blue dot-dashed line in Fig.~\ref{figmesonlims}. Notice that, for $m_h>2m_\mu$, this constraint results in a non-trivial excluded region in $(m_h,\sin^2\rho)$ parameter space. This is because the invisible yield can fall either by decreasing $\sin^2\rho$, thereby making the total cross section smaller, or by increasing $\sin^2\rho$, thereby making the decay more prompt. 

\begin{figure}
 \includegraphics[width=8cm]{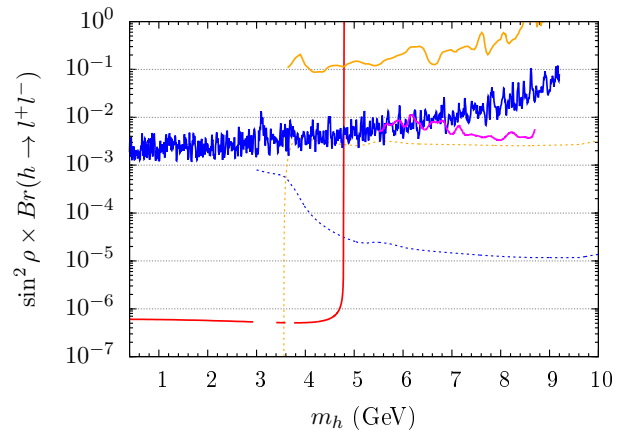}
 \caption{Upper limits on $\sin^2\rho\times Br(h\to l^+l^-)$ as a function of $m_h$ for $m_h>400$~MeV, where $l$ corresponds to either $\mu$ or $\tau$ depending on the channel: $B\to K\mu^+\mu^-$ (red), $\Upsilon\to \gamma h\to \gamma\mu^+\mu^-$ (blue), $\Upsilon\to \gamma h\to \gamma\tau^+\tau^-$ (orange), and $pp\to h \to \mu^+\mu^-$ via gluon fusion at CMS (magenta). Also shown is the level that dimuon (blue dashed) or ditau (orange dashed) bounds must reach to compete with L3 assuming branching ratios given by the perturbative approach in Sec.~\ref{secproperties}.}
 \label{figmesonlims2}
\end{figure}

\subsubsection{B meson decays}

The LHCb collaboration has measured $Br(B^+\rightarrow K^+\mu^+\mu^-)=(4.36\pm0.15\pm0.18)\times10^{-7}$ \cite{Aaij:2012vr}, the most accurate measurement to date and in good agreement with the theoretical prediction of $(3.5\pm1.2)\times10^{-7}$ \cite{Ali:2002jg}. However, we will use the results from B-factories \cite{Wei:2009zv,*Aubert:2008ps,Chen:2007zk,*delAmoSanchez:2010bk}, since the nature of an $e^+e^-$ collider makes it easier to predict the boost factor, and it is convenient to use the same experiment to constrain both the prompt and long-lived case:
\begin{align}
 Br(&B^+\rightarrow K^+ \mu^+\mu^-) \nonumber \\
 & =
 \begin{cases}
  (5.3^{+0.8}_{-0.7}\pm0.3)\times10^{-7} &\mbox{ (Belle)} \\
  (4.1^{+1.6}_{-1.5}\pm0.2)\times10^{-7} &\mbox{ (BaBar)}
 \end{cases}, \\
 & \approx (5.0\pm 0.8)\times10^{-7} \mbox{ (combined)}\\
 Br(&B^+\rightarrow K^+ \nu\bar{\nu}) \nonumber \\
 & <
 \begin{cases}
  1.4\times10^{-5} &\mbox{ (Belle)} \\
  1.3\times10^{-5} &\mbox{ (BaBar)}
 \end{cases},
\end{align}
where the combined visible decay bound is obtained by first adding statistical and systematic uncertainties for each measurement in quadrature and then combining the measurements in the usual way assuming they are independent unbiased estimators of $Br(B^+\rightarrow K^+ \mu^+\mu^-)$. A conservative limit on additive new physics is obtained by taking the difference between the low end of SM theoretical predictions, $Br(B^+\rightarrow K^+ \mu^+\mu^-)\gtrsim2.3\times10^{-7}_{theory}$, and the experimental measurement:
\begin{align}
 Br&(B\rightarrow K h) \times Br(h\rightarrow \mu^+\mu^-) \nonumber \\
 &\times \int_{0}^{\pi}\frac{\sin\theta d\theta}{2}
 \left(1-\exp\left[ \frac{-l_{xy}}{\sin\theta} \frac{1}{\gamma\beta c\tau} \right] \right)
 \lesssim 3\times 10^{-7}, \label{eqBKhmumu} \\
 Br&(B\rightarrow K h) \nonumber \\
 &\times \int_{0}^{\pi}\frac{\sin\theta d\theta}{2}
 \exp\left[ \frac{-l_{xy}}{\sin\theta} \frac{1}{\gamma\beta c\tau} \right]
 < 1.4\times 10^{-5},
\end{align}
where we follow Ref.~\cite{Batell:2009jf} in taking $l_{xy}\approx 25$~cm as the maximum reconstructed transverse decay distance from the beampipe, and $\gamma\beta\approx m_B/(2m_h)$ is dominated by the energy inherited from the B decay in the region $m_h<400$~MeV. The resulting bounds are shown in red in Figs.~\ref{figmesonlims} and \ref{figmesonlims2}. We do not set limits in the invariant mass regions surrounding $J/\psi$ and $\psi'$ since the experiments vetoed such muons to remove $B\to J/\psi X, \psi'X\to \mu^+\mu^-X$ background.

The visible B meson decay bound is stronger than the kaon bound since $K\to \pi h$ is CKM-suppressed compared to $B\to K h$. In the invisible case this suppression is overcome by the $\mathcal{O}(10^{-4})$ stronger bound resulting from a dedicated two-body kaon decay search. These bounds are enough to exclude the Foot \& Kobakhidze model for 100~MeV~$<m_h<(m_B-m_K)\approx 4.8$~GeV.

Visible decay bounds could be stronger if dedicated searches in the dimuon invariant mass spectrum were performed. Such a search was carried out in the $B^0\to K^{\ast0}X, (K^{\ast0}\to K^+\pi^-,X\to \mu^+\mu^-)$ channel at Belle in the region $212$~MeV$<m_X<300$~MeV \cite{Hyun:2010an}. No excess was found and an upper limit on the branching ratio of $\mathcal{O}(10^{-8})$ was set. Using this upper limit and the expression for $Br(B\rightarrow K^{\ast}h)$ in Ref.~\cite{Batell:2009jf} we derive a limit similarly to Eq.~\ref{eqBKhmumu}. This limit is given by the magenta line in Fig.~\ref{figmesonlims}.

LHCb could conceivably improve on the visible B decay bound, though we note that, as bounds reach below the $\sin^2\rho<10^{-5}$ level, special attention needs to be paid to $h$ lifetime. From Fig.~\ref{figbrctau} the mean decay length for $h$ in the region above the $\pi\pi$ threshold ranges between $10^{-9}$~cm~$/\sin^2\rho$ and $10^{-5}$~cm~$/\sin^2\rho$. These mean decay lengths are to be compared with those for B mesons, $c\tau_B\approx 5\times10^{-2}$~cm, for which LHCb measures displaced vertices. The lifetime of $h$ is of greater concern at LHCb where much larger boost factors are expected than at B-factories. We therefore encourage, as did Refs.~\cite{Freytsis:2009ct,Batell:2009jf,Bezrukov:2013fca}, a dedicated search for prompt decays covering the whole of the $\mu^+\mu^-$ invariant mass range, but we also recommend a displaced search. We will discuss this possibility further in the Sec.~\ref{seclhc}.

We note in passing that our benchmark particle cannot explain the $\Sigma^+ \rightarrow p\mu^+\mu^-$ HyperCP anomaly at $m_{\mu\mu} \approx 214$~MeV \cite{Park:2005eka}. Using the $\Sigma \to p h$ width in Ref.~\cite{Gorbunov:2005nu} we find that to match the measured branching fraction we require $\sin^2\rho \approx 10^{-5}$--$10^{-4}$. This region is disfavoured by visible B decays and invisible kaon decays. Even so, the lifetime of such a particle along with the expected boost factor of 100--200 suggested by the hyperon momentum gives $\gamma\beta c\tau > 13$~m, much larger than the longitudinal vertex resolution of about $0.2$~m \cite{Burnstein:2004uk}. Additionally, it appears that no value of $\mathcal{B}_{SM}$ could resolve the anomaly.

\subsubsection{Upsilon decays}

Limits also arise from the radiative $\Upsilon(nS)\to \gamma h$ decay shown in Fig.~\ref{figmesondecays}. The BaBar collaboration has searched in this channel for light bosons decaying to $\mu^+\mu^-$, $\tau^+\tau^-$, hadrons or escaping invisibly \cite{Lees:2012iw,Lees:2012te,Lees:2013vuj,delAmoSanchez:2010ac}. We reproduce the limits from dimuon and ditau decays \cite{Lees:2012iw,Lees:2012te} in Fig.~\ref{figmesonlims2} in solid blue and solid orange respectively, assuming the QCD correction factor $\mathcal{F}_{QCD}$ discussed therein is equal to unity. Ref.~\cite{McKeen:2008gd} discusses limits in light of CLEO data; for masses $m_h<2m_\tau$, scalar decays to pions and kaons can be more constraining than decays to muons (see Figure 14 of \cite{McKeen:2008gd}), though one must keep in mind the significant uncertainties in branching fractions.

B meson decays are easily more constraining for $m_h\lesssim(m_B-m_K)\approx4.8$~GeV. For $m_h\gtrsim4.8$~GeV, ditau limits give the best bound on $\sin^2\rho$ since $Br(h\to \tau^+\tau^-)$ is about $m_\tau^2/m_\mu^2\approx 287$ times larger than $Br(h\to \mu^+\mu^-)$. Even so, as can be seen from the dashed blue and dashed orange lines in Fig.~\ref{figmesonlims2}, these bounds do not yet challenge the L3 limit of $\sin^2\rho\lesssim 10^{-2}$.

\subsection{Fixed Target \label{subsecfixed}}

Our scalar $h$ can be produced either directly (through gluon fusion) or indirectly (via meson decays) in fixed target experiments. The dominant process depends on $\sqrt{s}$ and $m_h$. Meson decays dominate in the experiment we will consider below. 

Two important regions of parameter space may be identified for indirect production: below the kaon threshold, $m_h<(m_K-m_\pi)\approx360$~MeV, where kaon decays dominate, and below the B meson threshold, 360~MeV~$\lesssim m_h\lesssim m_B$, where B meson decays dominate. We note that there is a small region where $\eta$ decays can be important, but D meson decays are sufficiently CKM-suppressed to ignore. Some discussion and analysis may be found in Ref.~\cite{Bezrukov:2009yw}.

As an example, following Ref.~\cite{Bezrukov:2009yw}, we look at the bounds set by the CHARM Collaboration \cite{Bergsma:1985qz}. In this experiment, a 400~GeV proton beam was dumped into a thick copper target ($\sqrt{s}\approx \sqrt{2E_pm_p} \approx 27.4$~GeV) and the decay of a long-lived axion to photons, electrons or muons was searched for in a 35~m long decay region placed 480~m from the target. Zero decays were observed.

The total number of scalars intersecting the solid angle covered by the detector, $N_h$, is related to the number of decays in the decay region, $N_{dec}$, by
\begin{align}
 N_{dec} \approx & N_h \times \left[ Br(h\to e^+e^-)+Br(h\to\mu^+\mu^-) \right] \nonumber \\
 & \times \left[ -\exp\left(\frac{-L_2}{\gamma\beta c\tau}\right) + \exp\left(\frac{-L_1}{\gamma\beta c\tau}\right) \right],
\end{align}
where $\gamma\beta m_h\sim 10$~GeV, $L_1=L_2-35$~m $=480$~m, and $N_h\approx 2.9\times10^{17}\times \sigma_h/\sigma_{\pi_0}$ is normalised to the neutral pion yield \cite{Bergsma:1985qz}. We adopt $\sigma_{\pi_0}\approx \sigma_{pp} M_{pp}/3$, where $M_{pp}$ is the average hadron multiplicity and $\sigma_{pp}$ is the proton-proton cross section  \cite{Bezrukov:2009yw}. The $h$ production cross section is dominated by kaon decays:
\begin{align}
 \sigma_h\approx \sigma_{pp}M_{pp} \left[ 
\begin{array}{c}
 \chi_s\times \frac{1}{2} Br(K^+\to \pi^+ h) \\
 + \chi_s\times \frac{1}{4} Br(K_L\to \pi^0 h) \\
 + \chi_b\times Br(B\to h+X)
\end{array}
 \right],
\end{align}
where $\chi_s=1/7$, $\chi_b=3\times10^{-8}$, $Br(K_L\to \pi^0 h)=Br(K^+\to \pi^+ h)\times \Gamma(K^+)/\Gamma(K_L)$, and
\cite{Grinstein:1988yu}
\begin{align}
 Br(B\to h+X) \approx \sin^2\rho \times 0.26\left(\frac{m_t}{m_W}\right)^4\left(1-\frac{m_h^2}{m_B^2}\right)^2 . \label{eqBrBhX}
\end{align}

Since the CHARM experiment observed zero decays, we may constrain $N_{dec}$ at $90\%$ C.L. to be less than 2.3 (the solution of $0.1=\lambda^k e^{-\lambda}/k! \vert_{k=0}$). Our result is shown in Fig.~\ref{figmesonlims} by the green curve, with the enclosed region being excluded.
Observe that scalar masses 100~MeV~$<m_h<$~280~MeV are ruled out for the Bezrukov \& Gorbunov model by this analysis; the $K\to\pi+invisible$ and CHARM bounds also extend this exclusion substantially below 100~MeV, although it is not shown in Fig.~\ref{figmesonlims}.

The reach of the CHARM experiment is testament to the enormous production cross section of mesons in hadron collisions, as well as the exploitation of the long $h$ lifetime to remove all background. These two points, as we will see, are important for LHC phenomenology when $m_h\lesssim m_B$.

Other beam dump experiments exist which may complement the CHARM bound due to, in particular, differing beam energy and detector position \cite[for a partial list see Ref.][]{CooperSarkar:1985nh,*Badier:1986xz,*Bernardi:1987ek,*PhysRevD.52.6,*Adams:1997ht}. These include fixed target neutrino experiments, which have recently been considered as possibilities to probe GeV-scale portals \cite[see e.g.][]{Batell:2009di,*Essig:2010gu,*deNiverville:2011it,*deNiverville:2012ij}. It is beyond the scope of this paper to analyse these experiments in detail. However, we note that it does not appear that any of these experiments has probed the area above the eta meson threshold for $h$, because of insufficient direct or indirect production at given $\sqrt{s}$ (see Figure~30 of Ref.~\cite{Lourenco:2006vw} for B meson production rates) and/or the distance to the detector being too great. Ideally, high luminosity (and acceptance) fixed target experiments with energy $\sqrt{s}\gtrsim 20$~GeV and a detector placed at a 
distance $\mathcal{O}$(1--10~m) would be needed to probe parameter space below the B decay bound for $m_h\gtrsim 360$~MeV.

\section{LHC \label{seclhc}}

The $H\to hh$ channel may be phenomenologically relevant in models with a very light scalar. If allowed, this channel could produce back-to-back pairs of (possibly displaced) dimuons, for which searches have been carried out by ATLAS/CMS \cite{Aad:2012kw,*Aad:2012qua,*Chatrchyan:2011hr}, or contribute to the Higgs invisible width if $\mathcal{B}_{SM} < 1$. However, the effective couplings $Hhh$ and $Hhhh$ are independent of the parameters $m_h$ and $\sin^2\rho$, i.e. $H\to hh$ decay is not necessarily related to the scenario of a very light scalar mixing with the SM Higgs. For example, in the Foot \& Kobakhidze model, $Hhh$ and $Hhhh$ effective couplings are suppressed \cite{Foot:2011et}. Since we are focused on mixing-induced effects paramerised by $(m_h,\sin^2\rho)$ we do not consider this channel further.

\begin{table}
\begin{tabular}{c|c|c|c}
 \hline\hline
 Parton-level & \multicolumn{2}{|c|}{$\sigma(pp\to h+X)$ (pb)} & $\alpha_{ideal}$ \\
 process & $\sqrt{s}=7$~TeV & $\sqrt{s}=13$~TeV & $\sqrt{s}=7$~TeV \\
 \hline
 $gg\to h$ 		& $\sim 770$ 	& $\sim 1250$ 	& $\sim 5\times 10^{-4}$ \\
 $Wh$ 			& 170 		&  356		& $1.7\times 10^{-3}$ \\
 $Zh$ 			& 70 		&  147		& $2.3\times 10^{-3}$ \\
 $t\bar{t}h$		& 5.5 		&  27		& $2.4\times 10^{-2}$ \\
 $qq'h$			& 0.87 		&  1.9		& $1.3\times 10^{-1}$ \\
 \hline\hline
\end{tabular}
\caption{Parton-level cross sections contributing to $h$ production at the LHC for $m_h=5$~GeV, $\sin^2\rho=1$. Also shown is the acceptance factor $\alpha_{ideal}$ for a CMS dimuon search (see text). }
\label{tabxsecs}
\end{table}

For $m_h\lesssim m_B$, the dominant $h$ production mechanism at the LHC is via the production and decay of mesons. The $B\bar{B}$ cross section in 7~TeV (8/13~TeV) $pp$ collisions has been calculated as $\approx 2.5 \times 10^{11}$~fb ($\approx 3/10 \times 10^{11}$~fb) \cite{Lourenco:2006vw}. Then, for example, using Eq.~\ref{eqBrBhX}, at $\sin^2\rho=10^{-6}$ and $\sqrt{s}=7/8$~TeV the $h$ production cross section is $\sim 10^6$~fb, to be compared with $\sim1$~fb through gluon fusion. This is also an area of parameter space where $h$ lifetime becomes non-negligible. In the following subsection we determine the differential $p_T$ spectrum for scalars originating from B mesons at ATLAS/CMS and LHCb. We then show that this will result in up to thousands of moderate (triggerable) $p_T$ displaced decays in unexplored parameter space using the existing dataset. We note that for $m_h<m_K$ we also expect production via kaon decays. We ignore this area since, in our benchmark model, it has been explored by CHARM (see 
Sec.~\ref{subsecfixed}). Below the CHARM limit the lifetime becomes long enough so that the majority of moderate $p_T$ scalars would escape the detector. The situation may be different in models with $\mathcal{B}_{SM}<1$, since the lifetime becomes shorter, though one must take into account non-negligible kaon lifetime.	

For $m_h\gtrsim m_B$, $h$ is dominantly produced in the ways made familiar by the SM Higgs: gluon fusion, vector boson fusion, $Vh$, and $t\bar{t}h$. Table~\ref{tabxsecs} shows the production cross sections for an example scalar of mass 5~GeV and $\sin^2\rho=1$. Cross sections were obtained using the HiggsEffective model in MadGraph/MadEvent5 v1.5.9 \cite{Alwall:2011uj} equipped with CTEQ6L1 parton distribution functions \cite{Pumplin:2002vw}, except in the case of gluon fusion where we used \cite{Gunion:1989we}
\begin{align}
 \frac{d\sigma}{dy}\left(pp\to h\right) = &\frac{\pi^2}{8m_h^3} \Gamma(h\to gg) 
    \times g_p\left( \frac{m_h e^y}   {\sqrt{s}},m_h^2 \right)  \nonumber \\
   &\times g_p\left( \frac{m_h e^{-y}}{\sqrt{s}},m_h^2 \right) ,
   \label{eqggh}
\end{align}
where $g_p(x,Q^2)$ is the gluon distribution function in the proton evaluated at momentum fraction $x$ and scale $Q^2$, and we integrated over all possible rapidities $y$ using CTEQ5M parton distribution functions \cite{Lai:1999wy}.\footnote{MadGraph/MadEvent5 returns a value for gluon fusion of 670~pb in the $\sqrt{s}=7$~TeV case, but breaks at $\sqrt{s}=13$~TeV.} Gluon fusion is dominant, but $Vh$ production is comparable. Such associated production is important from an experimental point of view; trigger limitations and backgrounds affect the gluon fusion channel much more than for $Vh$ or $t\bar{t}h$. In the final subsection we demonstrate that the $Vh$ channel is in fact the most sensitive search at the LHC for $m_h\gtrsim m_B$.

\subsection{$\mathbf{m_h\lesssim m_B}$}

\subsubsection{Production via B decays}

\begin{figure}
 \includegraphics[width=8cm]{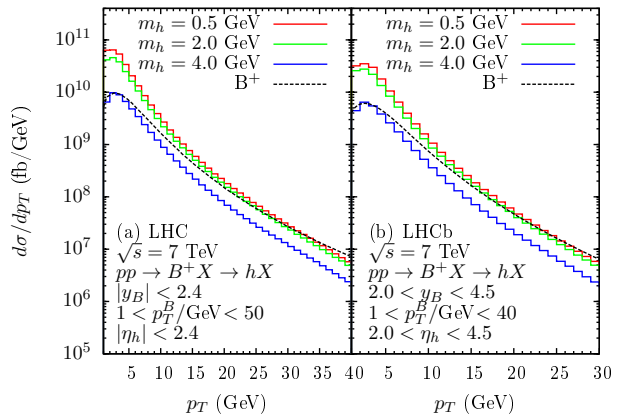}
 \caption{The differential cross section of $h$ production from $B^+$ decays in $\sqrt{s}=7$~TeV $pp$ collisions for $\sin^2\rho=1$ and $m_h=0.5,2.0,4.0$~GeV descending. The mother $B^+$ mesons are constrained in transverse momentum and rapidity as indicated.}
 \label{figBdecays}
\end{figure}

We developed an in-house simulation to calculate the differential cross section $d\sigma_h/dp_T$ for scalars from B decays, given $d\sigma_B/dp_T$ and $d\sigma_B/dy$ for B mesons in $\sqrt{s}=7$~TeV $pp$ collisions at ATLAS/CMS and at LHCb. It works in the following way: within loops over $p_T^B$ and $y_B$, there is a loop simulating $N_{dec}$ isotropic B decays to $h$, which are then boosted from the B frame to the lab frame given $p_T^B$ and $y_B$, rejected to a rejection bin in a histogram if they fall outside the angular acceptance, or else $p_T^h$ is measured and we add $f(p_T^B)f(y_B)/N_{dec}$ to the appropriate $p_T$ bin in a histogram,
where $f(p_T^B)$ and $f(y)$ define the discrete probability distributions for the transverse momentum and rapidity of the B meson. The histogram (which should now have unit area) is then normalised to $Br(B\to h+X)\times \int \frac{d\sigma_B}{dp_T} dp_T$. We infer $f(p_T^B)$ from the fixed-order-next-to-leading-logarithm (FONLL) predictions in Refs.~\cite{Cacciari:2012ny,ATLAS:2013cia,Aaij:2013noa}. This amounts to creating a probability density function by normalising $d\sigma_B/dp_T$ to unity over a chosen $p_T$ range and then discretising to allow for numerical integration. The $d\sigma_B/dp_T$ distributions used are reproduced in Fig.~\ref{figBdecays}. We interpolate $f(y)$ for ATLAS/CMS from the FONLL prediction in Figure~6 of Ref.~\cite{ATLAS:2013cia}, and for LHCb from the experimental measurements in Figure~4 of Ref.~\cite{Aaij:2013noa}. We make the approximations that $f(y)$ is independent of $p_T$, 
$|\vec{p}_h|$ in the B frame is equal to that from $B\to Kh$ decay, and the decay of the B meson is prompt.

Our results are shown in Fig.~\ref{figBdecays}. Note that we have only considered $B^+$ decays; results for $B^-$ would be identical, and for $B^0$/$\overline{B}^0$ would be very similar, so that the total $h$ cross section from B meson decay gains a factor $\approx 4$. For larger $m_h$ the $p_T$ tail falls more slowly because $h$ is produced at lower velocity in the B frame and therefore tends to follow the direction of the B meson. The overall cross section also falls due to kinematic suppression in $Br(B\to h+X)$.

With the information that is available to us, we are limited to using B mesons within a certain $p_T$ range and within rapidities that would be accepted at ATLAS/CMS or at LHCb. These limitations are written in Fig.~\ref{figBdecays} for clarity. Consequently, values of $d\sigma/dp_T$ in the LHCb case for $p_T\lesssim m_B$ are an underestimate, since smaller rapidity $B$ mesons will contribute. Otherwise we believe our results are a very good approximation. Ideally, one would loop over the entire range of allowed B momentum and rapidity using a complete $d^2\sigma_B/dp_T^Bdy_B$ prediction. 

The point to be gleaned from the distributions in Fig.~\ref{figBdecays} is that in unexplored parameter space with $\sin^2\rho<10^{-5}$ there are still a large number of moderate (triggerable) $p_T$ scalars being produced via B decay at ATLAS/CMS and at LHCb. Prompt $h$ decays will be best probed at LHCb by a ``bump search'' in the invariant dimuon mass of $B\to K\mu^+\mu^-$ decays, as explained in Sec.~\ref{subsecmesons}. For prompt decays ATLAS/CMS can only rely on inclusive dimuons, for which background at $m_{\mu\mu}<5$~GeV is large \cite[see e.g.][]{Chatrchyan:2011hr}. However, B-factory bounds are already pushing the boundary of non-negligible $h$ lifetime, which introduces the possibility of displaced decays as a way of removing background. For $m_h$ between 360~MeV and 5~GeV, $c\tau$ ranges between $10^{-9}$~cm~$/\sin^2\rho$ and $10^{-5}$~cm~$/\sin^2\rho$, to be compared with $\approx 5\times10^{-2}$~cm for B mesons which produce measurably displaced vertices at the LHC. Therefore, we expect a 
substantial region of parameter space with $\sin^2\rho<10^{-5}$ for which this model predicts many low-background displaced decays. It is this possibility that we pursue presently.

\subsubsection{Displaced decays}

The precedent for searches for displaced decays of light particles has already been set. ATLAS has performed a search for approximately back-to-back collimated dimuons originating from a 400~MeV particle decaying outside the inner detector but within the muon spectrometer, i.e. with transverse distance from the beamline 1~m~$\lesssim L_{xy}\lesssim$~7~m \cite{Aad:2012kw}. Prompt muon background is heavily suppressed -- there is almost zero background -- by requiring a lack of tracks in the inner detector within a cone surrounding the direction of the muon jet. No events are observed in 1.9~fb$^{-1}$ of data at $\sqrt{s}=7$~TeV. 

\begin{figure}
 \includegraphics[width=8cm]{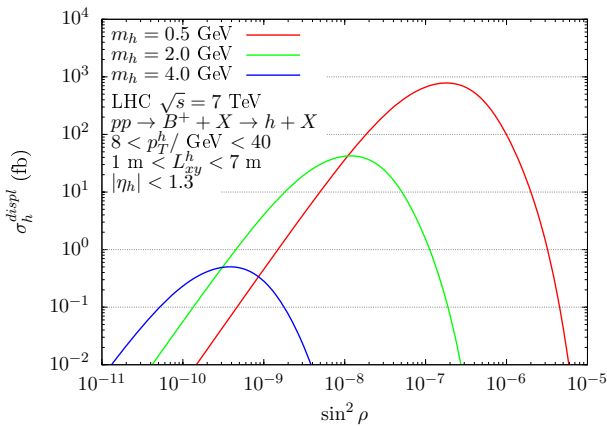}
 \caption{Cross section of moderate $p_T$ displaced (1~m~$\lesssim L_{xy}\lesssim$~7~m) decays of scalars originating from $B^+$ mesons produced in $\sqrt{s}=7$~TeV $pp$ collisions as a function of $\sin^2\rho$ for $m_h=0.5,2.0,4.0$~GeV descending.}
 \label{figdisplaced3}
\end{figure}

Such a search might be applied to $h$ to probe $H\to hh$ decays. However, motivated by the above analysis, we will consider the signature of an inclusive displaced muon pair. If we require the decay of $h$ to occur within transverse distance 1~m~$<L_{xy}<$~7~m then we expect a very low background.

Making the approximation $p \approx E$ ($\beta \approx 1$), the probability that a particle of mass $m$ will decay with transverse distance $L_1<L_{xy}<L_2$ from the beamline is given by
\begin{align}
 \mathcal{P}_{dec} \approx
 -\exp\left( \frac{-m L_2}{p_T \times c\tau} \right)
 +\exp\left( \frac{-m L_1}{p_T \times c\tau} \right) .
 \label{eqndecprob}
\end{align}
Note here that $c\tau$ is inversely proportional to $\sin^2\rho$ as in Eq.~\ref{eqctau}. As discussed in Sec.~\ref{secproperties}, the lifetime and branching fractions in the $2m_\pi<m_h<m_B$ region have a large uncertainty. With this in mind, in the following we evaluate $c\tau_{SM}$ using Ref.~\cite{Donoghue:1990xh} for $m_h<1.4$~GeV and the perturbative approach of Eq.~\ref{eqperturbative} otherwise. 

\begin{figure}
 \includegraphics[width=8cm]{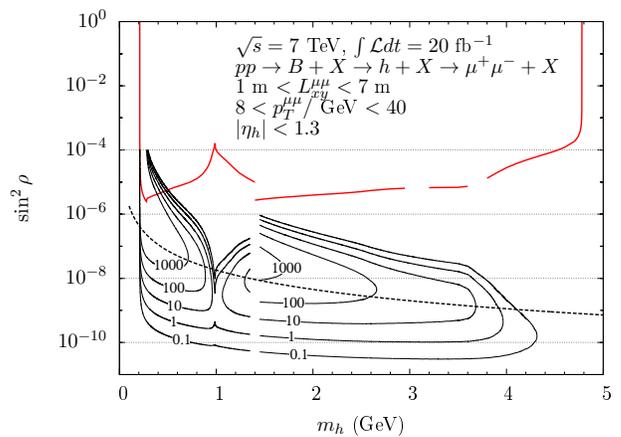}
 \caption{Contours representing a conservative underestimate of $0.1,1,10,100,1000$ moderate $p_T$ displaced dimuon decays of scalars originating from B mesons and occurring within the detector volume in 20~fb$^{-1}$ of $\sqrt{s}=7$~TeV data at the LHC (see text). Efficiency factors have not been considered. Shown above the contours in red is the bound from B decays obtained in Sec.~\ref{subsecmesons}. The discontinuity at $m_h=1.4$~GeV is due to uncertainties discussed in Sec.~\ref{secproperties}; we take the result of Ref.~\cite{Donoghue:1990xh} below $m_h=1.4$~GeV and the perturbative approach of Eq.~\ref{eqperturbative} otherwise. Also shown is the Bezrukov \& Gorbunov prediction (dashed).}
 \label{figdisplaced3-2}
\end{figure}

To estimate the reach of ATLAS/CMS we require the $h$ (dimuon) transverse momentum to satisfy $p_T^{h}>8$~GeV.
The cross section of displaced $h$ decays with $|\eta_h|<2.4$ can then be obtained from Fig.~\ref{figBdecays} by
 \begin{align}
  \sigma^{displ}_h \approx \sin^2\rho \times \int_{8\text{ GeV}}^{40\text{ GeV}} \mathcal{P}_{dec}(p_T) \frac{d\sigma_h}{dp_T} dp_T .
  \label{eqnsigmadec}
\end{align}
This will give a slight overestimate (by no more than a factor of 2) for the number of possibly observable displaced decays, since the requirements $|\eta_h|<2.4$ and 1~m~$\lesssim L_{xy}\lesssim$~7~m are not enough to ensure that the decay occurs within the detector volume. We therefore err on the conservative side by restricting $h$ to the central region $|\eta_h|<1.3$. This also ensures that the muons are created before the Level-1 muon trigger at ATLAS. Consequently, our results are an underestimate of the number of decays occuring within the detector volume.
The cross sections for three example masses are shown in Fig.~\ref{figdisplaced3}. As $\sin^2\rho$ gets smaller, the tuning of the mean decay length to $c\tau\sim$~1--100~cm to maximise $\mathcal{P}_{dec}$ plays off against the falling cross section to create a window in unexplored parameter space where the number of displaced decays can be significant.

If one wishes to search for displaced dimuons then the cross section in Fig.~\ref{figdisplaced3} must be scaled by $4\times Br(h\to \mu^+\mu^-)$. In Fig.~\ref{figdisplaced3-2} we show contours of the number of expected displaced dimuon events in 20~fb$^{-1}$ of data at $\sqrt{s}=7$~TeV; this plot serves to indicate the reach of the ATLAS/CMS 8~TeV dataset, which, in the absence of a $d\sigma_B/dp_T$ distribution for $\sqrt{s}=8$~TeV $pp$ collisions, we cannot generate the corresponding figure for. 

For example, at $m_h\approx 500$~MeV and $\sin^2\rho \approx 10^{-7}$, we predict (before efficiency factors) greater than $4\times10^3$ displaced collimated dimuons with $p_T^{\mu\mu}>8$~GeV. This scenario is consistent with the prediction of Bezrukov \& Gorbunov, shown in Fig.~\ref{figdisplaced3-2} as a dashed line. Notice that the area of parameter space which ATLAS/CMS is most sensitive to coincides with this line, meaning that the model may be extensively probed for scalar masses above the existing $m_h>280$~MeV limit.

In principle, a similar search could be performed at LHCb, although there exists no precedent. In fact, sensitivity is likely to be even better for a few reasons: smaller dimuon transverse momenta ($p_T^{\mu\mu}\approx1$~GeV) may be probed, the muon detection system extends to 19~m beyond the interaction point, and the vertex locater, with excellent reconstruction capabilities, might allow for probing of decays closer than 1~m. One drawback however is less integrated luminosity.

A dedicated study incorporating proper acceptance, trigger/reconstruction efficiency and backgrounds is desirable to say more about the reach of the LHC. We have required $h$ to fall within the central region, but this does not guarantee that each muon will have $|\eta_\mu|<2.4$. At least for lower $h$ masses, where the decay products will be collimated, this is a good assumption. Lower $m_h$ is also where we expect the efficiency to be highest, since efficiency falls with muon impact parameter and here the collimated muons will point back along the $h$ direction to the B decay point.  
SM backgrounds can only arise from neutral particles with lifetimes in the range $c\tau\sim$~1--100~cm. Of note are $K^0_S$ mesons ($c\tau_{K_S^0}\approx2.7$~cm) decaying to pions which may fake muons with $m_{\mu\mu}\approx500$~MeV either through decays-in-flight or punching through the calorimeters; such background appears to be well modelled by Monte-Carlo \cite{ATLAS-CONF-2010-064}. Neutral strange baryons $\Xi^0$ and $\Lambda^0$ with masses 1.3~GeV and 1.1~GeV respectively are the only other neutral SM particles with lifetimes in this range; it is not obvious how their decays could fake a $\mu^+\mu^-$ vertex. Therefore, at least for $m_h\gtrsim 500$~MeV, the background is expected to be very low so that even a few events, particularly since they will occur at the same dimuon invariant mass, may be significant. If necessary, SM background can always be suppressed by requiring $h$ to decay outside the hadronic calorimeter, 3~m~$\lesssim L_{xy}\lesssim$~7~m. In this regime one could also consider 
complementary signatures of decays to 
charged objects such as hadrons or $\tau^+\tau^-$ that might be picked up by the muon spectrometer. Further analysis is beyond the scope of this paper.

We have shown that with the existing dataset the LHC can (modulo efficiency factors) explore new parameter space by searching for displaced dimuons. Ultimately, knowledge of the exact excluded parameter space region is limited by the uncertainties in lifetime and branching fractions described in Sec.~\ref{secproperties}; we therefore encourage theorists to revisit that problem.

\subsection{$\mathbf{m_h\gtrsim m_B}$}

\subsubsection{Inclusive dimuon search}

Both ATLAS and CMS have performed a search for a light pseudoscalar, $a$, produced via gluon fusion and decaying to two muons \cite{ATLAS:2011cea, Chatrchyan:2012am}. The CMS search analysed the mass range between 5.5 and 8.8~GeV and between 11.5 and 14~GeV, avoiding the $\Upsilon$ resonances. They provide a $95\%$ C.L. upper limit on $\sigma(pp\to a)\times Br(a\to \mu^+\mu^-)$.

The production cross section of $h$ through the gluon fusion mechanism is given by Eq.~\ref{eqggh}. To constrain $h$ we assume that the acceptance in the CMS analysis is the same for our scalar as for the pseudoscalar, and consider only the dominant production of $h$ by gluon fusion. We then apply the $\sigma(pp\to a)\times Br(a\to \mu^+\mu^-)$ limit, evaluating Eq.~\ref{eqggh} by integrating over all possible rapidities using CTEQ5M parton distribution functions \cite{Lai:1999wy}. The result is the magenta line shown in Fig.~\ref{figmesonlims2}. The limit on $\sin^2\rho \times Br(h\to \mu^+\mu^-)$ competes well with that from upsilon decays, but is far from that of LEP.

The 1.3~fb$^{-1}$ of data analysed by CMS was collected with the opposite-sign dimuon trigger, requiring $p_T^{\mu\mu}>6$~GeV and $m_{\mu\mu}>5.5$~GeV with a prescale factor of 2. These low $p_T$, low invariant mass dimuons are evidently plentiful at the LHC. Thus, as the luminosity and centre-of-mass energy are increased the trigger thresholds and/or the prescale factor must increase. In short, we are background-restricted and trigger-restricted in the region that maximises signal.

So what happens if we demand high dimuon $p_T$, so as to minimise background and avoid trigger-dependence? CMS have performed a search for light resonances in the dimuon spectrum with 35~pb$^{-1}$ of data collected at $\sqrt{s}=7$~TeV \cite{Chatrchyan:2011hr}. At $m_h=5$~GeV, they set a $95\%$ C.L. limit on $\alpha_{ideal}\times \sigma(pp\to h+X)\times Br(h\to \mu^+\mu^-)<0.1$~pb, where $\alpha_{ideal}$ is an acceptance factor calculated in your favourite event generator by requiring
\begin{align}
 p_T^\mu &> 15 \text{ GeV}, \nonumber\\
 p_T^{\mu\mu} &>80 \text{ GeV}, \nonumber\\
 |\eta_\mu| &<0.9.
\end{align}
Using MadGraph/MadEvent5 we found $\alpha_{ideal}\approx 1.1\times 10^{-3}$ for $m_h=5$~GeV; it is broken down by channel in Table~\ref{tabxsecs}. For the gluon fusion channel we simulated $gg\to gh$ at parton-level, the hard gluon being necessary to give $h$ necessary $p_T$. Interestingly, every channel contributes comparable amounts to the result of $\alpha_{ideal}\times \sigma(pp\to h+X)\approx 1\times \sin^2\rho$~pb. From this we can constrain $\sin^2\rho\times Br(h\to \mu^+\mu^-)<0.1$ for $m_h=5$~GeV. Assuming that the bound will scale as $\sim 1/\sqrt{N}$, with 100 times more data -- comparable in size to the CMS pseudoscalar search -- we expect a bound of $\mathcal{O}(10^{-2})$. Therefore we have not gained anything on the pseudoscalar search bound by requiring high dimuon $p_T$. This is not surprising, since both the background and the dominant gluon fusion production mechanism have muons recoiling only against initial-state radiation, so that acceptance falls quickly with $p_T^{\mu\mu}$; 
this is reflected by the small value of $\alpha_{ideal}$ for gluon fusion in Table~\ref{tabxsecs}.


This leads us to consider instead triggering on associated activity so that some background is removed and we may probe lower $p_T$ muons from the $h$ decay. In the next section, we demonstrate that bounds using the $Wh$ channel, triggering on a high $p_T$ lepton from the $W$ decay, are potentially stronger than the bounds obtained from an inclusive dimuon search.

\subsubsection{Associated search}

There are three associated search possibilities: $Wh$, $Zh$, and $t\bar{t}h$. In this section we consider the $Wh\to W\mu^+\mu^-$ channel. Because it is in general difficult (and not just for us) to model the combinatoric background, we appeal to the results of experiment. ATLAS has performed a search in 4.6~fb$^{-1}$ of $\sqrt{s}=7$~TeV data for $J/\psi$ mesons produced in association with a $W$ boson, where both decay muonically \cite{ATLAS-CONF-2013-042}. The search amounts to a measurement of the ``bump size'' in the dimuon invariant mass spectrum around the $J/\psi$ mass of 3.1~GeV; they search in the region 2.5~GeV~$<m_{\mu\mu}<$~3.5~GeV. If $h$ exists in this region we would expect to see a bump above the combinatoric background. We aim to estimate the reach of a $Wh\to \mu\nu\mu^+\mu^-$ search using the background distribution therein.

We generate $Wh$ ($W\to\mu\nu$, $h\to\mu^+\mu^-$) parton-level events in $\sqrt{s}=7$ TeV $pp$ collisions for a scalar of mass 2.7~GeV with SM Higgs couplings using the HiggsEffective model in MadGraph/MadEvent5. We performed the following cuts to match those in Ref.~\cite{ATLAS-CONF-2013-042}:
\begin{align}
 |\eta_\mu| &<2.4 ,		& p_T^{\mu[1]} &>25\text{ GeV} , \nonumber\\
 \Delta R_{\mu\mu} &>0.3 ,	& p_T^{\mu[2]} &>4\text{ GeV}, \nonumber\\
 \slashed{E}_T &>20\text{ GeV}, &
 p_T^{\mu[3]} &>
 \begin{cases}
  3.5\text{ GeV} &\mbox{if } |\eta_{\mu[3]}|<1.3\\
  2.5\text{ GeV} &\mbox{if } |\eta_{\mu[3]}|>1.3
 \end{cases} ,
\end{align}
where the muons are ordered by $p_T$. We subsequently performed the following intermediate state cuts (which made little difference):
\begin{align}
  8.5\text{ GeV} &< p_T^h < 30\text{ GeV}, \nonumber\\
 |\eta_h| &<2.1 .
\end{align}
The results allow us to estimate the number of signal events in 4.6~fb$^{-1}$ of data as $\approx 1\times 10^{4}\times \sin^2\rho\times Br(h\to \mu^+\mu^-)$.

We take the combinatoric background and the number of observed events from Figure~2 of Ref.~\cite{ATLAS-CONF-2013-042}, restricting ourselves to the regions 2.50~GeV~$<m_{\mu\mu}<$~2.94~GeV and 3.28~GeV~$<m_{\mu\mu}<$~3.50~GeV to avoid the $J/\psi$ peak, since the peak is fitted to the data in this region. The signal is modelled as a gaussian with width 50~MeV and mean $m_h$.

\begin{figure}
 \includegraphics[width=8cm]{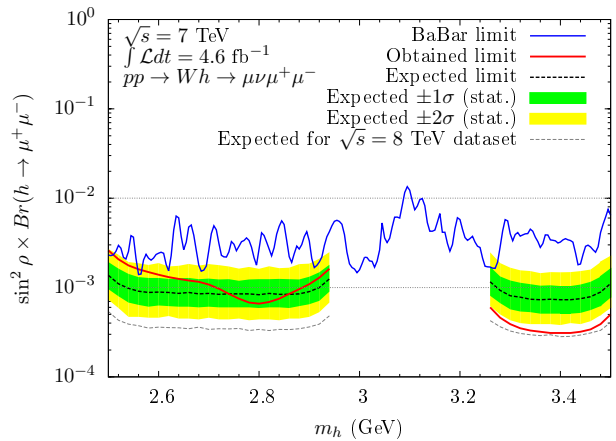}
 \caption{The obtained and expected $90\%$ C.L. upper limit on $\sin^2\rho\times Br(h\to\mu^+\mu^-)$ from the $Wh$ channel using 4.6~fb$^{-1}$ of $\sqrt{s}=7$~TeV data from ATLAS. Variance of the expected limit is statistical only. Also shown is an approximation of the expected limit using the 8~TeV dataset (see text) and the limit from $\Upsilon\to\gamma h\to\gamma\mu^+\mu^-$ decays.}
 \label{figwhstats}
\end{figure}

Let $\mu^b$ and $\mu^s$ be the vectors representing the expected number of background events and the expected number of signal events in $k$ bins. Let $y$ be the data vector. If we normalise $\mu^s$ to one event, then $\lambda\mu^s$ represents a signal bump with $\lambda$ total events. The likelihood of the data is
\begin{align}
 L(y|\lambda)=\prod_{j=1}^{k}(\mu^b_j+\lambda\mu^s_j)^{y_j}\exp\left[ -(\mu^b_j+\lambda\mu^s_j) \right].
\end{align}
Bayes' theorem relates this likelihood to our degree of belief in $\lambda$:
\begin{align}
 p(\lambda|y) \propto L(y|\lambda)\pi(\lambda),
\end{align}
where $\pi$ is the prior distribution for $\lambda$. If we take a flat prior, 
\begin{align}
 \pi(\lambda) =
 \begin{cases}
  1 & \lambda \geq 0, \\
  0 & \mbox{otherwise},
 \end{cases}
\end{align}
then the $90\%$ C.L. upper limit on $\lambda$, $\lambda_{UL}$, is found by solving
\begin{align}
 0.90 = \int_0^{\lambda_{UL}} \hat{L}(y|\lambda) ,
\end{align}
where $\hat{L}$ has been normalised so that $\int_0^\infty \hat{L}(y|\lambda)=1$. The $90\%$ C.L. upper limit on $\sin^2\rho\times Br(h\to \mu^+\mu^-)$ is then simply $10^{-4}\lambda_{UL}$. We have performed this analysis for a signal centred on each of 36 $m_h$ values spread 20~MeV apart.

The obtained upper limit is given by the red line in Fig.~\ref{figwhstats}. An expected $(\pm1\sigma/\pm2\sigma$ stat.$)$ limit was derived by performing the above analysis on $10^3$ pseudodatasets generated assuming the background only hypothesis, ordering them by the obtained $\lambda_{UL}$, and taking entry 500 $(^{841}_{159}/^{977}_{023})$, shown by the dashed line and bands in Fig.~\ref{figwhstats}. We also show the expected limit for the case with five times the data, which serves as an approximation for the reach of the 8~TeV dataset. One can see that the limit of $\mathcal{O}(10^{-3})$ is better than that set by radiative upsilon decays. A similar limit would be expected for $m_h> m_B$, potentially setting the best LHC limit on $\sin^2\rho \times Br(h\to \mu^+\mu^-)$ in that region. However, as is evident from Fig.~\ref{figmesonlims2}, it would still be two orders of magnitude weaker than the L3 limit.

We note that the expected sensitivity of a $Zh$ search, where both the $Z$ and $h$ decay muonically, is expected to be higher because the extra lepton would help to remove combinatoric background. In the future, a search for the production of prompt $J/\psi$ mesons in association with a $Z$ boson may allow the above analysis to be reperformed. The reach of the 13~TeV run is not clear because we do not know the combinatoric background, but one could speculate that more data and higher sensitivity in the $Zh$ channel may be enough to compete with LEP bounds of $\mathcal{O}(10^{-5})$.

\section{Conclusion \label{secconclusion}}

Motivated by scale invariant and inflationary models, we investigated the phenomenology of a very light scalar, $h$, with mass 100~MeV~$<m_h<$~10~GeV, mixing with the SM Higgs. As a benchmark model we took the real singlet scalar extension of the SM and explored ($m_h,\sin^2\rho$) parameter space, where $\rho$ is the mixing angle. 

The existing limits are shown in Figs.~\ref{figLEPlims}, \ref{figmesonlims} and \ref{figmesonlims2}. 
For $100$~MeV~$<m_h<210$~MeV, fixed target experiments and $B\to K+invisible$ decays limit $\sin^2\rho \lesssim 10^{-8}$. 
For $210$~MeV~$<m_h<280$~MeV, $B\to K\mu^+\mu^-$ decays and fixed target experiments rule out almost all of the parameter space above $\sin^2\rho=10^{-10}$ (see Fig.~\ref{figmesonlims}).
For $280$~MeV~$<m_h<360$~MeV, the same experiments constrain $\sin^2\rho \lesssim 10^{-10}$ except for a window between $10^{-8}\lesssim \sin^2\rho \lesssim 10^{-5}$ which is still allowed.
For $360$~MeV~$<m_h<4.8$~GeV, $B\to K\mu^+\mu^-$ decays limit $\sin^2\rho\times Br(h\to \mu^+\mu^-) \lesssim 10^{-6}$.
For $4.8$~GeV~$<m_h<10$~GeV, searches for the Bjorken process $Z\to Z^\ast h\to Z^\ast+ hadrons$ at LEP1 give the best limit of $\sin^2\rho \lesssim 10^{-2}$.

At the LHC we identified two phenomenologically distinct regions of parameter space. For $m_h\lesssim m_B$, $h$ is dominantly produced via the decay of B mesons, with a rate $\sim10^6$ times larger than gluon fusion. In regions of parameter space where $h$ decays promptly, $\sin^2\rho \gtrsim 10^{-5}$, LHCb could set the best limits by searching for resonances in the $B\to K\mu^+\mu^-$ dimuon invariant mass spectrum. In the region $\sin^2\rho \lesssim 10^{-5}$, $h$ lifetime is non-negligible. We investigated the possibility of searching for displaced dimuons at ATLAS/CMS, showing that, in unexplored parameter space coinciding with the model of Bezrukoz \& Gorbunov \cite{Bezrukov:2009yw}, more than $10^3$ signal events (before efficiency factors) could be in the existing 8~TeV dataset (see Fig.~\ref{figdisplaced3-2}). By requiring the muons to exhibit no track in the inner detector we expect this search to be almost background-free. This motivates a search for inclusive displaced dimuons at 
ATLAS/CMS and/or LHCb.

For $m_h\gtrsim m_B$ we demonstrated that the subdominant $Vh$ production channel has the best sensitivity at ATLAS/CMS. Bounds from the $Wh$ channel using 4.6~fb$^{-1}$ of $\sqrt{s}=7$~TeV data were found to be $\sin^2\rho\times Br(h\to\mu^+\mu^-)\lesssim 10^{-3}$ in the region 2.5~GeV~$<m_h<3.5$~GeV (see Fig.~\ref{figwhstats}). This limit is stronger than that from upsilon decays, and is expected to extend into the $m_h>m_B$ region if the analysis was performed. Such a bound would still be about two orders of magnitude weaker than that of LEP1. We expect that the $Zh$ channel would provide better sensitivity and it is conceivable that future LHC bounds could compete with that of LEP1, with the main uncertainty being knowledge of the combinatoric background at $\sqrt{s}=13$~TeV.

In Sec.~\ref{secproperties} we highlighted apparently unresolved uncertainties in the branching ratios and lifetime of $h$ in the region 280~MeV~$<m_h\lesssim4$~GeV. This is the region that exhibits interesting displaced LHC phenomenology. The most recent paper dedicated to this subject, that we are aware of, is over twenty years old; we therefore recommend that the theory community revisit the problem.

Lastly, we note that a similar analysis could be performed for a very light scalar mixing with the SM Higgs and also decaying to hidden states, by appropriately scaling parameters as described in the Introduction.

\addvspace{1cm}
\noindent\textbf{Note added:} After completion of this paper, Ref.~\cite{Schmidt-Hoberg:2013hba} appeared on the arXiv, which has some overlap with Sec.~\ref{subsecmesons}.

\begin{acknowledgments}

This work was supported in part by the Australian Research Council.
JDC would like to thank Tony Limosani for experimental input on the $t\bar{t}h$ channel (that unfortunately did not end up appearing in this paper) and the suggestion of using Ref.~\cite{ATLAS-CONF-2013-042} to estimate background in the $Vh$ channel, as well as Evgueni Goudzovski for input on the NA48/2 kaon decay bound. 

\end{acknowledgments}

\bibliography{references}

\end{document}